\documentclass[runningheads]{llncs}
\usepackage[T1]{fontenc}
\usepackage{graphicx}
\usepackage{float}
\usepackage{verbatim}
\usepackage{xcolor}
\usepackage{graphicx}
\usepackage{geometry}
\usepackage{multicol}
\usepackage{enumitem} 
\usepackage{amssymb}
\usepackage{array}
\usepackage{longtable}
\usepackage{caption}
\usepackage{subcaption}
\usepackage{wrapfig}
\usepackage{fancyvrb}
\usepackage{bigstrut}
\usepackage[english]{babel}
\usepackage{pgfplots}
\usepackage{mdwlist}
\usepackage{multirow} 
\usepackage{url}
\usepackage{boxedminipage}
\usepackage{verbatim}
\usepackage{flafter}
\usepackage{listings}
\usepackage[figuresright]{rotating}
\usepackage{epstopdf} 
\usepackage{placeins}
\usepackage{comment}
\usepackage{color}
\usepackage{tabularx}
\usepackage{listings}
\usepackage{float}
\usepackage{hyperref}
\usepackage{listings}
\usepackage{url}
\usepackage{listings}
\usepackage{multirow}

\newcommand{\REM}[1]{}
\usepackage{float}
\usepackage{hyperref}
\usepackage{booktabs}
\usepackage{multirow}
\newsavebox{\codebox}
\usepackage{subcaption}

\usepackage{comment}
\usepackage{booktabs}
\usepackage{color}

\usepackage{booktabs}
\usepackage{multirow}
\usepackage{listings}
\usepackage{backnaur}

\usepackage{amsmath,amssymb,amsfonts}
\usepackage{algorithmic}
\usepackage{graphicx}
\usepackage{textcomp}
\usepackage{xcolor}
\usepackage{verbatim}

\usepackage{listings}
\usepackage{comment}
\begin{document}
\title{Developing an Interactive OpenMP Programming Book with Large Language Models}
%
%
\author{Xinyao Yi\inst{1}\orcidID{0000-0002-6614-1274} \and
Anjia Wang\inst{2}\orcidID{0000-0001-7092-7408} \and
Yonghong Yan\inst{1}\orcidID{0000-0002-5274-8526} \and Chunhua Liao\inst{3}\orcidID{0000-0001-6477-0547}}
\authorrunning{X. Yi et al.}
\institute{University of North Carolina at Charlotte, Charlotte, NC, USA\\ 
\email{\{xyi2,yyan7\}@charlotte.edu} \and
Intel Corporation, Hillsboro, OR, USA\\
\email{anjia.wang@intel.com} \and
Lawrence Livermore National Laboratory, CA, USA\\
\email{liao6@llnl.gov}}
\maketitle              
\begin{abstract}
This paper presents an approach to authoring a textbook titled \texttt{Interactive OpenMP Programming} with the assistance of Large Language Models (LLMs). 
The writing process utilized state-of-the-art LLMs, including Gemini Pro 1.5, Claude 3, and ChatGPT-4, to generate the initial structure and outline of the book, as well as the initial content for specific chapters. This content included detailed descriptions of individual OpenMP constructs and practical programming examples.
The outline and content have then undergone extensive manual revisions to meet our book goals. In this paper, we report our findings about the capabilities and limitations of these LLMs. We address critical questions concerning the necessity of textbook resources and the effectiveness of LLMs in creating fundamental and practical programming content. Our findings suggest that while LLMs offer significant advantages in generating textbook content, they require careful integration with traditional educational methodologies to ensure depth, accuracy, and pedagogical effectiveness.
The \texttt{Interactive OpenMP Programming} book is developed with the framework of Jupyter Book, enabling the execution of code within the book from the web browser, providing instant feedback and a dynamic learning experience that stands in contrast to traditional educational resources.
The book represents a significant step towards modernizing programming education, offering insights into practical strategies for generating the textbook through advanced AI tools.


\end{abstract}
\section{Introduction}
\label{sec:intro}
Given the increasing complexity of supercomputer node architectures in high-performance computing (HPC), high-level programming models have become essential to enhance productivity. 
OpenMP is a critical programming model in parallel computing, widely used for multi-core, multi-threaded processors, many-core accelerator architectures, and a combination of them. 
While interest in utilizing OpenMP in HPC is growing, the OpenMP language has evolved to feature more complex syntax. The length of the OpenMP specification has expanded from 318 pages in OpenMP 3.1 to 649 pages in OpenMP 5.2, creating a steep learning curve for this high-level programming model. The OpenMP framework's complexity, stemming from its advanced execution models and the in-depth knowledge required to manage parallel tasks, poses significant educational challenges. 
Traditional educational resources, such as textbooks for OpenMP, often fail to incorporate the latest programming additions and lack the interactivity necessary for effective learning. Outdated content and limited accessibility frequently hinder learning efficiency.

Recently, the landscape of computing and application has been evolving with the rapid advancement of artificial intelligence technologies, particularly large language models (LLMs) such as Gemini Pro 1.5, Claude 3, and ChatGPT-4. LLMs provide a prompt-answer model of learning, offering interactive and personalized educational experiences without the need for an actual human teacher. These include grammar and syntax support, case-based learning, concept clarification, and customized exercises and quizzes. While LLMs can significantly ease the learning process for OpenMP, their utilization in this context presents unique challenges. LLMs demonstrate proficiency in handling specific tasks and adeptly addressing specific questions. However, they fall short of providing a comprehensive learning schedule covering all OpenMP aspects. In contrast to traditional textbooks, LLMs are less effective in guiding students through a structured and progressive learning process. Additionally, effective prompts are crucial to optimizing the accuracy and relevance of LLM-generated content.

To leverage the strengths of both LLMs and traditional resources while addressing their limitations, we propose the development of an interactive OpenMP book. This book aims to stay updated with the latest OpenMP advancements and implement an interactive learning experience. We utilize LLMs' powerful text generation capabilities to rapidly generate the initial version, including explanations of OpenMP constructs, programming examples, and code interpretations. Designed with a "learning by practice" approach, the book provides up-to-date code examples and allows learners to experiment with them immediately. 
The interactive book is open-sourced and available at \url{https://passlab.github.io/InteractiveOpenMPProgramming/cover.html}.

The main contributions of this work include:
\begin{itemize}
  \item Generating an interactive book named \texttt{Interactive OpenMP Programming};
  \item Proposing a method for rapidly generating OpenMP educational content using LLMs;
  \item Evaluating the effectiveness and limitations of LLMs in creating both conceptual and practical programming content;
  \item Demonstrating how well-designed prompts can significantly enhance the quality of content generated by LLMs, contributing to more accurate educational materials.
\end{itemize}

The rest of the paper is as follows: 
Section~\ref{sec:motivation} discusses the background and the motivation for using LLMs to develop the \texttt{Interactive OpenMP Programming} book and identifies the deficiencies of traditional textbooks. Section~\ref{sec:method} details the methodology employed in utilizing LLMs to create an interactive programming book, including the strategies for prompt design and content validation. 
Section~\ref{sec:evaluation} assesses and discusses the content generated by the LLMs and the content in \texttt{Interactive OpenMP Programming} book.
Section~\ref{sec:related_work} introduced the related work.
The paper concludes in Section~\ref{sec:conclusion}, where we summarize our findings, outline future research avenues, and suggest improvements for more effectively integrating LLMs into educational frameworks.

\section{Background and Motivation}
\label{sec:motivation}

Learning OpenMP programming through LLMs and traditional textbooks each has its own set of advantages and disadvantages. Table~\ref{tab:pros_cons_llm_textbook} provides a comparison of these methods.

LLMs offer several advantages over traditional textbooks. Primarily, they provide instant feedback and interactive engagement. They could provide immediate responses to inquiries, analyze user-submitted code in real time, and suggest enhancements. 
Additionally, LLMs are highly adaptable and customizable educational tools. They can tailor content and examples to match the user's skill level and learning preferences, potentially speeding up the learning process. Furthermore, LLMs can access various information sources, providing diverse perspectives and solutions. 
In contrast, traditional textbooks do not offer these advantages. They lack the interactivity essential for effective learning and are unable to adjust content in real time based on students' needs. 

\begin{table}[]
\centering
\begin{tabular}{|p{0.45\textwidth}|p{0.45\textwidth}|}
\hline
\multicolumn{2}{|c|}{\textbf{Learn from LLMs}} \\
\hline
\textbf{Pros} & \textbf{Cons} \\
\hline
    - Deliver instantaneous feedback and interactive engagement.
    
    - Provide immediate responses to inquiries.
    
    - Analyze user-submitted code in real-time and suggest enhancements.
    
    - Serve as highly adaptable and customizable educational tools.
    
    - Access a wide array of information sources, offering diverse perspectives and solutions.
&
    - Offer a relatively superficial depth of understanding of complex OpenMP constructs.
    
    - Unable to verify the accuracy of the information they generate.
    
    - Unable to execute the code in real-time.
    
    - Lack of a structured learning trajectory.
    
    - Pre-trained with fixed knowledge about a field, making it difficult to update with new information.
\\
\hline
\multicolumn{2}{|c|}{\textbf{Learn from Traditional Textbooks}} \\
\hline
\textbf{Pros} & \textbf{Cons} \\
\hline
    - Provide a systematic and comprehensive exploration of OpenMP programming, aligning with the principle of progressive learning.
    
    - Incorporate case studies and best practices that are the culmination of years of expert experience and scholarly research.
    
    - Uphold a high standard of information accuracy.
&
    - Lack of the essential interactivity needed for effective learning.
    
    - Struggle to keep pace with the latest programming paradigms and updates.
    
    - Require high costs for purchase, making some books less accessible.
\\
\hline
\end{tabular}
\caption{Advantages and Limitations of Learning from LLMs and Traditional Textbooks}
\label{tab:pros_cons_llm_textbook}
\vspace{-0.5cm}
\end{table}

Despite the advantages, the drawbacks of learning using LLMs cannot be ignored.
The first drawback is the relatively superficial depth of understanding. From our experience with SIMD and vector architecture in OpenMP, while LLMs can introduce the basic usage of \texttt{\#pragma omp simd}, they often fall short in providing detailed discussions on optimizing SIMD for specific hardware configurations. In contrast, textbooks are typically authored by experienced experts and scholars, ensuring a deeper understanding of the OpenMP content.
Another issue is the absence of mechanisms within LLMs to verify the accuracy of the information they generate, potentially leading to the dissemination of incorrect or misleading content. For example, LLMs can generate explanations for \texttt{\#pragma omp task depend}, 
but we noticed that they fail to handle data race conditions in multi-threading environments. In contrast, textbooks maintain a high standard of information accuracy.
Lastly,  LLMs lack structured and progressive education procedures.
Traditional textbooks typically provide a more systematic and comprehensive exploration of OpenMP programming, incorporating case studies and best practices. 
They follow pedagogical principles, building foundational knowledge before advancing to complex topics. Although LLMs can be programmed to offer structured education, their flexibility can lead to deviations. For example, traditional learning for GPU programming starts with basic architecture and progresses to advanced topics like memory management. LLMs, however, might skip essential steps or present advanced topics prematurely based on user queries, resulting in fragmented or incomplete understanding.

A common drawback of both LLMs and textbooks is their inability to stay updated with the latest OpenMP specifications. LLMs, being pre-trained, possess fixed knowledge about a field and are challenging to update with new information. For example, in the OpenMP specification 5.2, the \texttt{depend} clause is no longer used with the \texttt{ordered} directive. Despite attempts to update ChatGPT with the latest specifications, it still occasionally generates content that incorrectly uses the \texttt{depend} clause. We find that this issue can be mitigated by directing the LLMs to learn and use specific new information. However, beginners might not be aware of whether the content generated by an LLM complies with the latest specifications, nor understand how to prompt the LLMs to produce updated content. This problem is more pronounced and difficult to resolve in textbooks. Once published, textbooks are hard to modify, and we often find that many contain outdated content that cannot be promptly updated.

Based on these pros and cons, we aim to combine the best of both worlds by using LLMs to help develop a structured, interactive textbook - \texttt{Interactive OpenMP Programming}. This involves using advanced AI tools to modernize programming education and enhance its effectiveness, thereby providing learners with a more interactive, engaging, and effective educational experience. 
First, the decision to employ LLMs arises from the need to address the limitations of traditional OpenMP educational resources. LLMs can quickly generate up-to-date educational content, solving the problem of outdated material in traditional textbooks.
Secondly, we will analyze the depth of LLMs' understanding of different OpenMP constructs and manually complete sections where the LLM's understanding is insufficient. At the same time, we will rigorously review the accuracy of the content generated. This task is initially aimed at addressing issues with LLMs' depth of understanding of complex topics and the inability to audit the correctness of generated content. Furthermore, our exploration will provide a reference for others studying LLMs' comprehension of OpenMP.
Next, we will use Jupyter Notebooks to develop an interactive, progressive textbook, where learners can execute their code directly to verify its correctness. This approach maintains the structured learning benefits provided by traditional textbooks while addressing their limitations in supporting interactive learning, the inability of LLMs to offer systematic, progressive learning, and the lack of real-time code execution.
Finally, the motivation also includes exploring how to optimize LLM output through the strategic design of prompts, which is essential for enhancing the quality of the generated content. This approach ensures that the material is not only technically accurate and pedagogically sound but also tailored to the specific learning context of OpenMP programming.

\section{Method}
This study employs a multifaceted approach to address challenges in content generation for OpenMP programming using Large Language Models (LLMs). By leveraging the strengths of multiple LLMs, including Gemini Pro 1.5, Claude 3, and ChatGPT-4, we mitigate the limitations of individual models. A Chain-of-Thought method is applied to overcome the context window size limit, dividing the process into two stages: outline generation and content development. Prompt engineering is optimized using the CO-STAR framework, while one-shot learning ensures consistency by referencing prior chapters. Evolving OpenMP specifications are incorporated through in-context learning, enriching the generated content with official examples. Cross-model review and manual edits are employed to enhance quality to address hallucinations and inconsistencies. Finally, the generated content is integrated into Jupyter Notebooks, facilitating interactive code execution and immediate feedback for learners. This approach effectively combines AI-driven generation with manual refinement, ensuring accuracy and educational value.
\label{sec:method}

\subsection{Outlining the Book and Each Chapter with the Aid of LLMs}

A combined approach of human expertise and machine-generated content is recommended to effectively utilize LLMs in composing a comprehensive OpenMP programming book. The methodology involves the following steps:

\subsubsection{Generating Outlines of the Textbook}
Using various LLMs to create multiple book structures based on their interpretations of the OpenMP API 5.2 Specification and analyses of the official OpenMP API 5.2.2 Examples~\cite{ompexample} will leverage the LLMs' capability to quickly process and synthesize extensive datasets.
We used the prompt: "Here are the OpenMP 5.2 Specification and the official examples. Please use this information to guide the creation of the textbook outline. Include sections on basic and advanced topics, referencing specific constructs from the specification and relevant examples. Outline a few initial chapters focusing on core concepts, synchronization, and tasking".

However, we discovered that the frameworks produced by the LLMs did not align with our intended organizational approach. 
For instance, ChatGPT allocated substantial sections to discussing the foundational concepts of OpenMP and the setup processes. However, the detailed explanations and examples of the directives and the clauses were confined to merely two chapters-"Fundamentals of OpenMP" and "Advanced OpenMP". Similarly, the other two LLMs we evaluated did not perform satisfactorily. Consequently, after reviewing the frameworks generated by three different LLMs and integrating professional insights, we manually developed the book's structure. Our book emphasizes analyzing and applying various directives and clauses, illustrated through a combination of fundamental and advanced examples.  

\subsubsection{Generating Outlines for Chapters}
Although the LLMs offered limited assistance in establishing the overall framework, they proved highly beneficial in crafting detailed outlines and contents for individual chapters.
To ask the LLMs to generate the outline of the chapter "2.4. Synchronization of Threads Using Barrier and Ordered Directive," we first had the LLMs learn about synchronization in parallel computing scenarios from OpenMP Specification and official examples. Then, we uploaded a manually completed chapter on the \texttt{teams} construct to serve as a reference for the LLMs.
\begin{table}[]
\begin{tabular}{|p{0.12\textwidth}|p{0.3\textwidth}|p{0.55\textwidth}|}
\hline
\textbf{Guidelines} &
  \textbf{Explanation} &
  \textbf{Our Prompts} \\ \hline
Context(C) &
  Providing background information &
  I am currently \textbf{writing a book} on OpenMP parallel programming aimed at teaching others. \textbf{I have completed a chapter on teams}, which follows a specific outline style. I am now \textbf{focusing on the synchronization} of OpenMP, specifically on barrier and order constructs. \\ \hline
Objective (O) &
  Clearly defining the task &
  \textbf{Generate an outline} for the chapter on synchronization, focusing on barrier and order constructs. The outline should be based on the style used in the previous chapter on teams. \\ \hline
Style (S) &
  Specifying the writing style &
  Educational, structured \\ \hline
Tone (T) &
  Setting the tone &
  Instructional, clear \\ \hline
Audience (A) &
  Identifying the intended audience &
  Readers are students and programmers new to parallel programming, as well as educators looking for teaching resources. \\ \hline
Response (R) &
  Providing the response format &
  Provide \textbf{a structured outline} in text format that details sections and subsections for the chapter on synchronization, including key points to cover for barrier and order constructs. \\ \hline
\end{tabular}
\caption{The CO-STAR Framework and Its Example Use for Outlining the Chapter on Synchronization}
\label{tab:costaroutline}
\end{table}
To use the proper prompts, we incorporate the CO-STAR framework, which outlines a structured approach to crafting prompts that can significantly enhance the effectiveness and professionalism of the content generated by LLMs~\cite{costar}. Developed by GovTech Singapore’s Data Science and Artificial Intelligence Division, CO-STAR stands for Context, Objective, Style, Tone, Audience, and Response. The prompt guidelines, explanations, and examples are shown in Table~\ref{tab:costaroutline}.

The prompts are progressive and interactive, involving multiple steps, including learning from uploaded material, analyzing a previous chapter, and developing a new chapter outline.  This approach generates a more detailed and contextually enriched outline.
We've found that the structure generated by this set of prompts closely resembles that of the previous \texttt{teams} chapter and incorporates many terms like \texttt{explore}, \texttt{understanding}, \texttt{analysis}, and \texttt{apply} that are more apt for educational contexts.
The generated and revised outlines are demonstrated in Table~\ref{tab:outlines}.

\subsubsection{Assessment and Selection of the Outlines}
We followed a three-step assessment process to ensure the quality and accuracy of the generated outlines. First, we conducted a cross-model review, where one LLM critiqued another's outlines, enhancing them through mutual learning. Next, according to OpenMP Specification, we manually inspected each outline's comprehensiveness and accuracy. Finally, we integrated the structures. We merged the best aspects of each outline, balancing their strengths and weaknesses to create an optimized and effective structure.
    
\begin{table}[h!]\scriptsize
 \begin{tabular}{|p{0.5\textwidth}|p{0.5\textwidth}|}
    \hline 

 \begin{itemize}
 \item[] \textbf{Outline generated by Gemini Pro 1.5}
 \vspace{0.1 cm}
    \item \textbf{Introduction}
    
    \item \textbf{Barrier Directive}
    \begin{itemize}
        \item Establishing Synchronization Points
        \item Example
    \end{itemize}
    
    \item \textbf{The Ordered Directive: Maintaining Sequential Steps}
    \begin{itemize}
        \item Enforcing Order in the Parallel Ballet
        \item \texttt{doacross} Clause: Specifying Dependencies
    \end{itemize}
    
    \item \textbf{Implicit Barriers: Automatic Synchronization}
    \begin{itemize}
        \item Removing Implicit Barriers with \texttt{nowait}
    \end{itemize}
    
    \item \textbf{Best Practices for Using Barrier and Ordered Directives}
\end{itemize}

    &  
 \begin{itemize}
  \item[] \textbf{Outline generated by ChatGPT-4}
  \vspace{0.1 cm}
    \item \textbf{Introduction}
    \item \textbf{Barrier Directive}
    \begin{itemize}
        \item Purpose of the Barrier Directive
        \item Usage
        \item Example: Using the Barrier Directive
        \item Considerations
    \end{itemize}
    
    \item \textbf{Ordered Directive}
    \begin{itemize}
        \item Purpose of the Ordered Directive
        \item Usage
        \item Compatibility with the doacross Clause
        \item Example: Serial Output in Parallel Loop
        \item Considerations
    \end{itemize}
    
    \item \textbf{Summary}
    \begin{itemize}
        \item Key Takeaways
    \end{itemize}
\end{itemize}
\\
    \hline 
\begin{itemize}
\item[] \textbf{Outline generated by Claude 3}
\vspace{0.1 cm}
    \item \textbf{Introduction}
    \begin{itemize}
        \item Importance of Thread Synchronization
        \item Overview of the Barrier and Ordered Directives
    \end{itemize}
    \item \textbf{Barrier Directive}
    \begin{itemize}
        \item Purpose and Usage
        \item Syntax and Examples
        \item Barrier Regions
        \item Synchronization Points
    \end{itemize}
    \item \textbf{Ordered Directive}
    \begin{itemize}
        \item Purpose and Usage
        \item Syntax and Examples
        \item Enforcing Execution Order
        \item Ordered Regions
        \item Stand-alone Ordered Construct
        \item Block-associated Ordered Construct
        \item Interaction with Loop Constructs and Clauses
        \item Best Practices
    \end{itemize}
    \item \textbf{Combining Barrier and Ordered Directives}
    \begin{itemize}
        \item Use Cases for Combining Directives
        \item Examples and Code Snippets
        \item Considerations and Potential Issues
    \end{itemize}
    \item \textbf{Implicit Barriers}
    \begin{itemize}
        \item Implicit Barrier Regions
        \item Execution Model Events and Tool Callbacks
    \end{itemize}
    \item \textbf{Advanced Topics}
    \begin{itemize}
        \item Nested Barrier and Ordered Directives
        \item Interoperability with Other Synchronization Mechanisms
        \item Synchronization in the Context of Tasking
        \item Debugging and Profiling Synchronization Issues
    \end{itemize}
    \item \textbf{Performance Considerations}
    \begin{itemize}
        \item Overhead and Scalability
        \item Load Balancing and Synchronization Granularity
        \item Performance Tuning and Optimization
    \end{itemize}
    \item \textbf{Summary and Conclusion}
\end{itemize}
        &  
\begin{itemize}
  \item[] \textbf{Revised Outline}
  \vspace{0.1 cm}
    \item \textbf{Introduction}
    \begin{itemize}
        \item Importance of Thread Synchronization
        \item Overview of the Barrier and Ordered Directives
    \end{itemize}

    \item \textbf{Barrier Directive}
    \begin{itemize}
        \item Purpose and Usage
        \item Syntax and Practical Examples
        \item Barrier Regions and Synchronization Points
    \end{itemize}

    \item \textbf{Ordered Directive}
    \begin{itemize}
        \item Purpose and Usage
        \item Syntax and Practical Examples
        \item Interaction with Loop Constructs and \texttt{doacross} Clause
        \item Best Practices and Considerations
    \end{itemize}

    \item \textbf{Combining Barrier and Ordered Directives}
    \begin{itemize}
        \item Use Cases for Combining Directives
        \item Examples and Code Snippets
        \item Considerations and Potential Issues
    \end{itemize}

    \item \textbf{Implicit Barriers}
    \begin{itemize}
        \item Overview and Automatic Synchronization
        \item Controlling Implicit Barriers
    \end{itemize}

    \item \textbf{Advanced Topics}
    \begin{itemize}
        \item Nested Directives and Interoperability
        \item Synchronization in the Context of Tasking
        \item Debugging and Profiling Synchronization Issues
    \end{itemize}

    \item \textbf{Performance Considerations}
    \begin{itemize}
        \item Overhead and Scalability
        \item Load Balancing and Synchronization Granularity
    \end{itemize}

    \item \textbf{Summary and Conclusion}
    \begin{itemize}
        \item Recap of Key Points
        \item Further Learning and Applications
    \end{itemize}
\end{itemize}
    \\
    \hline 
\end{tabular}
    \caption{Outlines Generated by Gemini Pro 1.5, ChatGPT-4 and Claude 3, and the Revised Outline Used in the Textbook for Section 2.4. Synchronization of Threads Using Barrier and Ordered Directives}
    \label{tab:outlines}
\vspace{-1cm}
\end{table}

The outlines generated by different LLMs have distinct characteristics.
The outline generated by Gemini Pro 1.5 is concise, focusing on practical applications and key concepts. It briefly introduces the basic usage and examples of \texttt{barrier} and \texttt{ordered} directives, explaining how to set synchronization points and maintain execution order. However, Gemini lacks comprehensive detail in practical application examples and discussions of advanced topics.
The outline generated by ChatGPT 4 features clearly structured steps and examples, such as the basic usage of Barrier directives and compatibility with the \texttt{doacross} clause. This outline may offer some technical depth while maintaining good generality and overview. 
The outline generated by Claude 3 is a highly detailed structure covering basic concepts and uses while exploring various complex scenarios and performance considerations, such as teaching advanced uses of \texttt{barrier} and \texttt{ordered} directives like nested directives and interoperability with task scheduling. However, it might be overly complex, including some unnecessary details, especially for beginners, requiring a longer learning curve.

\subsection{Content Generation, Including Code Examples using Three LLMs}

\subsubsection{Understanding and Explaining Fundamental Constructs}

\begin{figure}[htb!]
\captionsetup{font=footnotesize}
\begin{minipage}{\linewidth}
\centering
\includegraphics[width=1.0\textwidth]{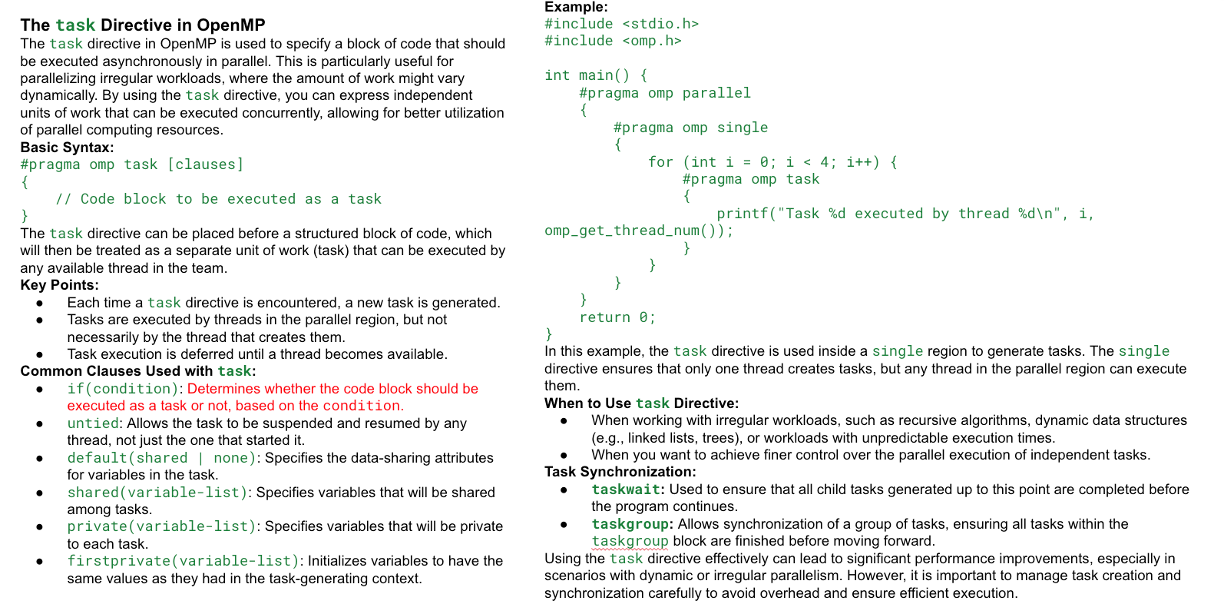}
\subcaption{}
\end{minipage}
\vfill
\begin{minipage}{\linewidth}
\centering
\includegraphics[width=1.0\textwidth]{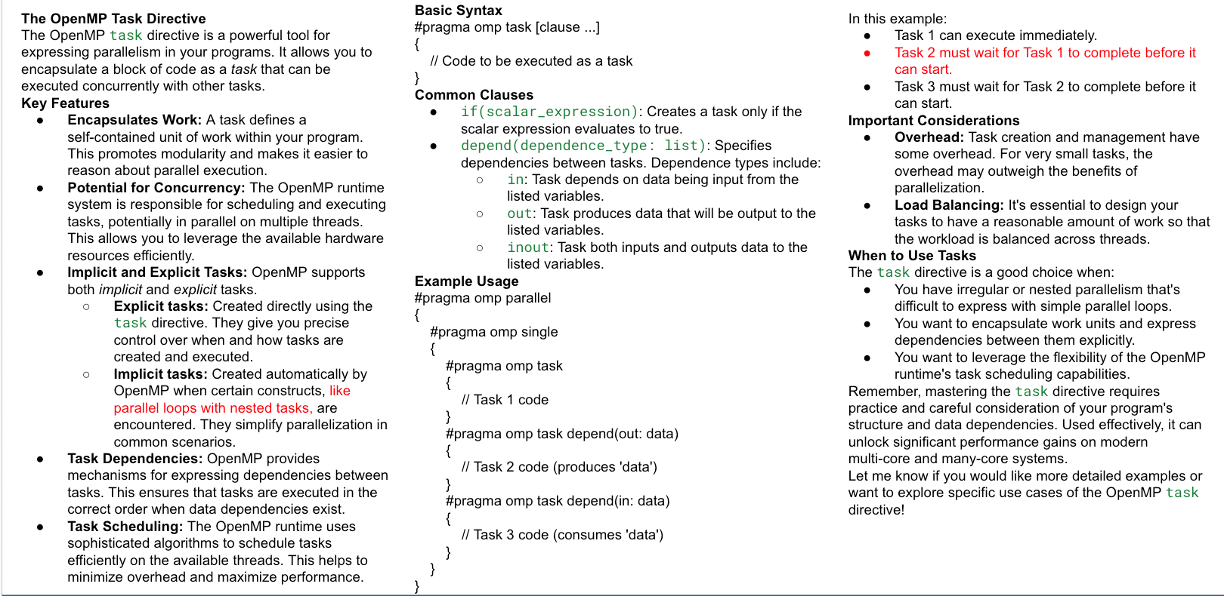}
\subcaption{}
\end{minipage}
\vspace{-0.3cm}
\caption{Descriptions of the \texttt{task} Directive Generated by Different LLMs}
\label{fig:TASKdescriptions}
\end{figure}

This subsection outlines our analysis of how LLMs generate specific text. We show the prompts used in Table~\ref{tab:task} and show the descriptions of the \texttt{task} directive generated by different LLMs in Figure~\ref{fig:TASKdescriptions}.

\begin{table}[]
\begin{tabular}{|p{0.18\textwidth}|p{0.8\textwidth}|}
\hline
\textbf{CO-STAR Guidelines} &
  \textbf{Our Prompts} 
   \\ \hline
Context(C) &
  The task directive is an essential component of OpenMP used to define independent units of work that can be executed in parallel. It's particularly useful for handling irregular workloads in parallel computing. 
   \\ \hline
Objective (O) &
  Provide a comprehensive explanation of the task directive, including its syntax, clauses, examples, and best practices for using it in OpenMP programs. 
   \\ \hline
Style (S) &
  Educational, detailed, and structured 
   \\ \hline
Tone (T) &
  Informative, encouraging, and supportive 
   \\ \hline
Audience (A) &
  Students, programmers, and practitioners who are learning OpenMP and wish to understand how to effectively utilize the task directive for parallel programming. 
   \\ \hline
Response (R) &
  Generate the content in a clear, step-by-step manner suitable for textbook material. Include an explanation of the task directive, examples with code snippets, and guidance on when and how to use it effectively. 
   \\ \hline
\end{tabular}
\caption{Prompts Used for Generating the Descriptions of \textbf{task} Directive}
\label{tab:task}
\end{table}

Here, we only present the results generated by ChatGPT and Gemini, as the content generated by Claude was too simplistic and lacked depth and detail, thus not included. ChatGPT's output provided a clear and concise explanation of the \texttt{task} directive in OpenMP, focusing on its use for parallelizing irregular workloads. The generated code example was simple and effective, demonstrating basic syntax and task creation. In contrast, Gemini generated a more detailed explanation, covering both explicit and implicit tasks, and introduced task dependencies using the \texttt{depend} clause. This additional depth makes it more suitable for beginner and intermediate readers.

However, we identified several errors in the content generated by both LLMs, which we have highlighted in red in the images. In ChatGPT’s output, the statement "if(condition): Determines whether the code block should be executed as a task or not, based on the condition" is inaccurate. In fact, when an \texttt{if} clause is present on a task construct and the expression evaluates to false, an undeferred task is generated, and the encountering thread must suspend the current task region. Execution cannot resume until the structured block associated with the generated task is completed. Additionally, using a variable in the \texttt{if} clause expression causes an implicit reference to the variable in all enclosing constructs. The evaluation of the \texttt{if} clause expression occurs outside the context of the task construct. 

In Gemini’s output, the statement "Implicit tasks: Created automatically by OpenMP when certain constructs, like parallel loops with nested tasks, are encountered" is incorrect. Implicit tasks are created by OpenMP parallel regions. Additionally, when explaining the generated example, the assertion that "task 2 must wait for task 1 to complete" is incorrect. Task 2 does not need to wait for task 1 unless explicit dependencies are defined, making the description inaccurate.

These examples highlight the need for a manual review of LLM-generated content. While LLMs can generate informative and accessible technical content, errors, oversimplifications, and inaccuracies can still occur. Therefore, human oversight is essential to ensure that the content is both technically accurate and complete.


All three LLMs accurately describe the basic functionality of the \texttt{barrier} directive for technical accuracy. 
Claude provides a straightforward introduction to the \texttt{barrier} directive's role in thread synchronization, focusing on its basic functionality. 
Gemini and ChatGPT provide a detailed and context-rich explanation, emphasizing the \texttt{barrier} directive's critical role in preventing race conditions, which could be very beneficial in understanding more complex scenarios.

Considering the educational purpose, Claude offers a more generalized and accessible introduction, making it suitable for beginners just starting to learn about thread synchronization in parallel programming. ChatGPT balances a general introduction and a specific application, making it ideal for intermediate learners. With its detailed explanations of particular scenarios like race conditions, Gemini targets readers with some background in parallel programming and looking to deepen their understanding of synchronization challenges.

\subsubsection{Generation and Explanation of Examples}

Different LLMs showcase unique difficulty, depth, and correctness attributes when analyzing the generated examples. The prompt we used to generate the examples is shown in Table~\ref{example_prompts}

\begin{table}[]
\begin{tabular}{|p{0.15\textwidth}|p{0.8\textwidth}|}
\hline
\textbf{CO-STAR Guidelines} & \textbf{Our Prompts}                                                             \\ \hline
Context(C) &
  I am currently writing a book on OpenMP parallel programming aimed at teaching others. In this chapter, we will focus on {[}directive name{]}. \\ \hline
Objective (O) &
  \textbf{Create an example} that demonstrates the use of a {[}directive name{]} directive, particularly illustrating the functionality of a named clause and how it can help manage parallelism. \\ \hline
Style (S)                   & Educational, detailed                                                            \\ \hline
Tone (T)                    & Instructional, clear                                                             \\ \hline
Audience (A) &
  Programmers and computer science students who are learning about thread synchronization and parallel processing. \\ \hline
Response (R)                & Provide a code snippet in C or C++ that uses the specified directive and clause. \\ \hline
\end{tabular}
\caption{Prompts Used for Generating Examples and Example Explanation}
\label{example_prompts}
\end{table}

\begin{figure}[!htb]
\captionsetup{font=footnotesize}
\begin{lrbox}{\codebox}%
\begin{minipage}[b]{0.3\linewidth}%
\begin{Verbatim}[fontsize=\scriptsize, frame=single, numbers=left, xleftmargin=0.03\linewidth]
#include <stdio.h>
#include <omp.h>
#define N 1000
int main() {
  float a[N], b[N], result[N];
  float factor = 0.5;
  // Initialize arrays
  for (int i = 0; i < N; i++){
    a[i] = i * 1.0f;
    b[i] = (N - i) * 1.0f;
  }
  // Apply SIMD directive with 
  // linear clause
  #pragma omp simd \
    linear(factor:0.1)
  for (int i = 0; i < N; i++) {
    result[i]=a[i]+b[i]*factor;
  }
  return 0;
}
\end{Verbatim}
\end{minipage}%
\end{lrbox}%
\subfloat[OpenMP SIMD Example Generated by ChatGPT]{\usebox{\codebox}}%
\hfill
\begin{lrbox}{\codebox}%
\begin{minipage}[b]{0.3\linewidth}%
\begin{Verbatim}[fontsize=\scriptsize, frame=single, numbers=left, xleftmargin=0.03\linewidth]
#include <stdio.h>

#define N 1024

int main() {
  float a[N], b[N], c[N];
  // Initialize arrays
  for (int i = 0; i < N; i++) {
    a[i] = i * 2.0f;
    b[i] = i * 3.0f;
  }
  
  // SIMD computation with 
  // preferred simdlen of 4
  #pragma omp simd simdlen(4)
  for (int i = 0; i < N; i++) {
    c[i] = a[i] + b[I];
  }
  return 0;
}
\end{Verbatim}
\end{minipage}%
\end{lrbox}%
\subfloat[OpenMP SIMD Example Generated by Gemini]{\usebox{\codebox}}%
\hfill
\begin{lrbox}{\codebox}%
\begin{minipage}[b]{0.3\linewidth}%
\begin{Verbatim}[fontsize=\scriptsize, frame=single, numbers=left, xleftmargin=0.03\linewidth]
#include <stdio.h>

int main() {
  float a[N], b[N], c[N];
  float sum = 0.0;
  int i;
  
  // Initialize the arrays
  // Do the sum
  #pragma omp simd
  for (i = 0; i < N; i++) {
    c[i] = a[i] + b[I];
  }
  #pragma omp parallel for \
    reduction(+:sum)
  for (i = 0; i < N; i++) {
    sum += c[I];
  }
  return 0;
}
\end{Verbatim}
\end{minipage}%
\end{lrbox}%
\subfloat[OpenMP SIMD Example Generated by Claude]{\usebox{\codebox}}%
\caption{Comparison of SIMD Implementations Generated by GPT-4, Gemini Pro 1.5, and Claude 3.}
\label{fig:example_compare}
\end{figure}

Figure~\ref{fig:example_compare} shows three examples of OpenMP SIMD code generated by GPT-4, Gemini Pro 1.5, and Claude 3 demonstrate different levels of detail and approach when handling SIMD directives in OpenMP. Each example reflects varying styles and technical nuances, which could affect their educational value and applicability in practice.

In the GPT-4 generated code, a basic OpenMP SIMD example is provided, where the \texttt{simd} directive is applied with a linear clause. The code initializes two arrays (\texttt{a[]} and \texttt{b[]}), and then performs element-wise multiplication, storing the results in \texttt{result[]}. The inclusion of the \texttt{linear(factor:0.1)} clause ensures that the \texttt{factor} variable is adjusted incrementally across iterations. This example is more sophisticated because it illustrates the use of a clause that modifies a variable within the loop, giving learners a more advanced understanding of how OpenMP SIMD can be applied with specific optimization directives. However, the SIMD directive applied here does not specify a preferred vector length, leaving the optimization choices to the compiler.

The code generated by Gemini Pro 1.5 uses a slightly different approach. The SIMD directive includes the \texttt{simdlen(4)} clause, explicitly specifying a preferred vector length for SIMD operations, indicating a more performance-optimized approach. The example similarly initializes arrays but focuses on element-wise addition (\texttt{c[] = a[] + b[]}). By specifying a vector length, this example provides a clear way to control the hardware-level parallelism, making it highly suitable for users interested in low-level performance tuning. The use of \texttt{simdlen(4)} also shows an important feature in SIMD that allows fine-tuning of vectorization to match the underlying hardware capabilities.

Claude 3’s generated example is the simplest of the three. The code initializes three arrays and performs a sum reduction using both SIMD and parallel reduction directives. The first part of the code uses the \texttt{simd} directive for the element-wise addition (\texttt{c[] = a[] + b[]}), followed by a parallelized sum reduction across the array \texttt{c[]}. This example demonstrates the combination of SIMD with another common parallel construct (\texttt{reduction}), which is useful for illustrating more complex parallel patterns. However, it lacks detailed optimization controls (such as the \texttt{simdlen} clause or any additional optimizations), making it more suitable for beginners or those focusing on basic syntax and functionality.

The GPT-4 and Gemini Pro 1.5 examples offer more advanced usage of SIMD compared to Claude 3, with GPT-4 focusing on linear variable adjustments and Gemini Pro 1.5 incorporating hardware-level optimizations through the \texttt{simdlen} clause. Claude 3's example is simpler, combining basic SIMD operations with a parallel reduction.
Gemini Pro 1.5 stands out in terms of performance optimization, providing explicit control over vector length, which can lead to better performance on specific hardware. GPT-4 uses the \texttt{linear} clause, which is educational for variable handling in SIMD but does not provide low-level performance tuning.
GPT-4’s example is valuable for understanding how clauses like \texttt{linear} can be used within SIMD loops, making it suitable for learners exploring variable handling in parallel loops. Gemini Pro 1.5’s example is ideal for those focusing on performance tuning with SIMD. Claude 3’s example, while simpler, effectively shows how SIMD can be integrated with parallel reduction, making it more approachable for beginners or as a first introduction to combined parallel constructs.

We conducted a detailed analysis of each generated example, noting that each LLM demonstrated varying levels of understanding for different directives. Overall, in terms of both generating and explaining examples, GPT-4 consistently performed the best, followed by Gemini.

\subsubsection{The Workflow of Generating Interactive Format with the Help from LLMs}
In this subsection, we will elaborate on generating and updating content for an interactive book with the help of LLMs, which we ultimately create and edit as JSON files in a Jupyter Notebook. Theoretically, it is possible to have LLMs produce text in a specific format since generating formatted text is a strength of LLMs. However, in practice, it is not as straightforward. One of the interactive features of our designed book allows users to enter their code directly into a code input box and execute it to receive feedback. This necessitates the setting up of separate code cells, specifying the programming language (C/C++ or Fortran) and the compiler. Initially, LLMs do not always generate the format we want, often requiring repetitive debugging and constant modification of our instructions to make the LLM understand the desired format, which is quite inefficient.
Moreover, different LLMs have varying levels of understanding and capability. For instance, ChatGPT can usually generate text in different formats, Gemini may not fulfill formatting requirements, and Claude can produce Markdown format but cannot separate code from textual descriptions. Therefore, we only have LLMs generate content, ideally in Markdown format, without insisting that they produce directly usable JSON files, which must be manually adjusted for format and cell type later.

Updating the \texttt{Interactive OpenMP Programming} book is straightforward. When changes are needed, we can update the content in the local repository and then upload these changes to the online repository using the standard GitHub pull request process.
The online GitHub repository automatically manages the book deployment. Typically, the book is compiled locally before uploading to ensure there are no compilation errors that could hinder proper deployment. Similarly, we allow users to add content to the book or upload their code. The difference is that users need to apply before uploading. Our administrators review the content and decide whether to merge it into our book.

\subsection{Interactive and Incremental Development of a Programming Book}

\begin{figure*}[!htb]
\centering
\includegraphics[width=0.7\linewidth]{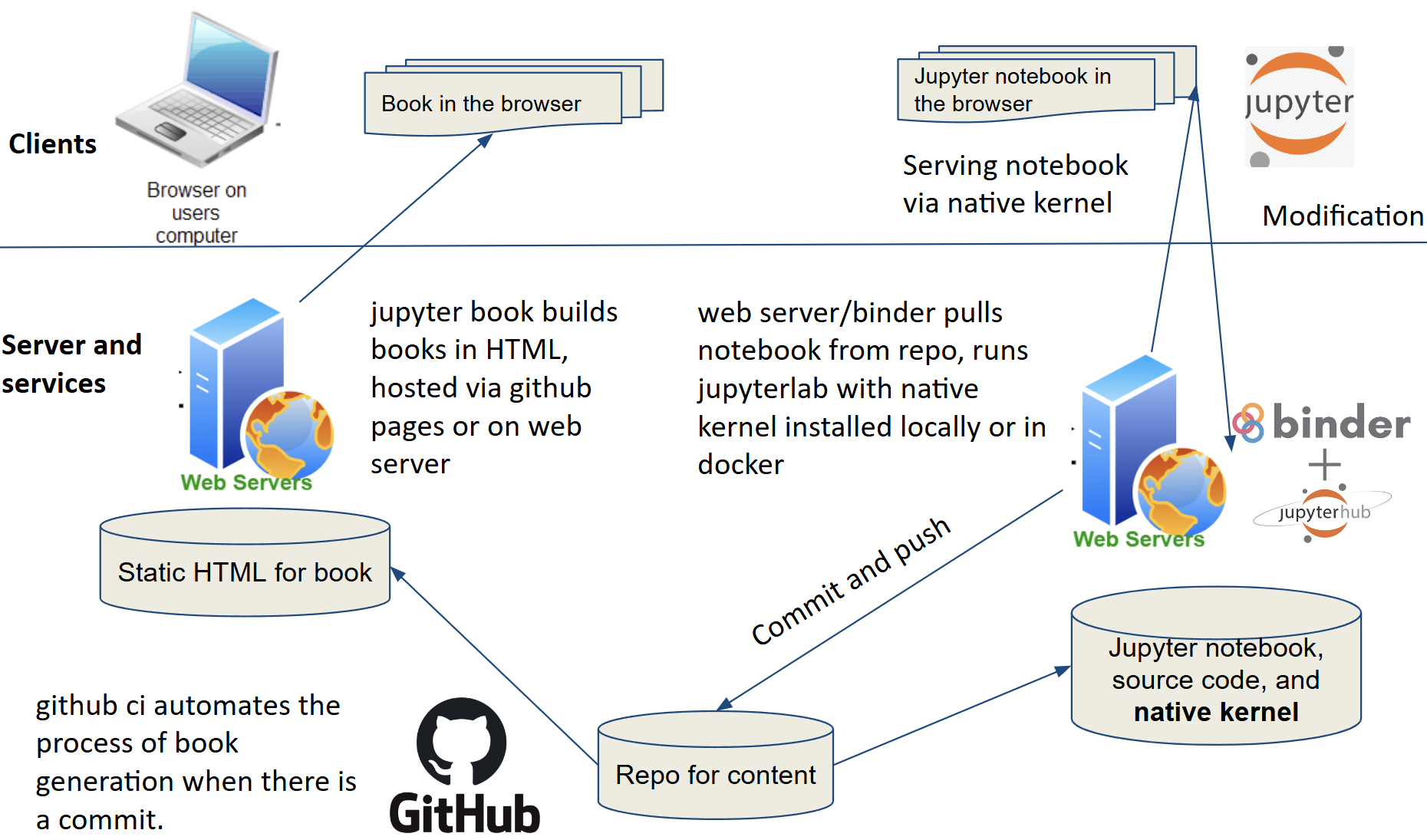}
\caption{Client and Server Architecture of Interactive OpenMP Programming Book}
\label{fig:book_arch}
\vspace{-0.5 cm}
\end{figure*}

The design of the interactive OpenMP programming book is shown in Figure~\ref{fig:book_arch}. On the client side, users can access the book from a browser on any web-enabled device.
Besides conventional reading instructions, they can modify the corresponding Jupyter notebooks and conduct experiments.
For the server side, all the book sources are stored on GitHub. This generates both the book and Jupyter notebooks. The reading materials are provided as HTML files~\cite{interactive_omp_book}. The Jupyter notebooks, which act as a coding sandbox, are served via JupyterLab with a native kernel.
There are three ways to deploy the book, each with strengths and limitations.

\subsubsection{Local Deployment}
We can deploy the interactive OpenMP programming book on a local Linux machine by installing JupyterBook and JupyterLab.
JupyterBook generates the book website in HTML.
JupyterLab runs the code examples in the book on demand. 
Local deployment is suitable for trial runs and development. The user can quickly modify the book content or backend server configuration and review the changes.

\subsubsection{Self-hosting Deployment}
Self-hosting is similar to local deployment in most aspects, except a server for the public or a specific private network is required.
The benefit of self-hosting deployment is that we can create a more capable programming sandbox for users. 
For example, a local laptop may not support OpenMP GPU offloading, while a self-hosting server does.
However, a self-hosting server requires more professional maintenance work. It typically must provide uninterrupted service.

\subsubsection{Third-party-hosting Deployment}
Another way to deploy the programming book is to utilize third-party services like Binder~\cite{mybinder}. Instead of setting up the service from scratch on a local or remote machine under control, we can create a Dockerfile to specify what OS and software should be installed.
Binder will create an online virtual machine based on the Dockerfile and set up the sandbox accordingly.
However, users can't control Binder server's hardware configuration, which is typically not very powerful.




\section{Assessment and Discussion}
\label{sec:evaluation}
\subsection{Comparing the Learning Capabilities of Different Models}
We have analyzed and evaluated the capability of various LLMs to learn from uploaded materials. The results are summarized in Table~\ref{tab:model_comparison}.

\begin{table}[]
\centering
\begin{tabular}{|p{0.25\textwidth}|p{0.23\textwidth}|p{0.23\textwidth}|p{0.23\textwidth}|}
\hline
\textbf{Aspect of}

\textbf{Evaluation}   & \textbf{ChatGPT}                  & \textbf{Gemini}                    & \textbf{Claude}                    \\ \hline
\textbf{Initial}

\textbf{Understanding}  & \textcolor{red}{\textbullet} Incorrect 

(used depend)   & \textcolor{green}{\textbullet} Correct (doacross 

clause) & \textcolor{red}{\textbullet} Failed (did not 

recognize) \\ \hline
\textbf{Comprehensiveness}      & \textcolor{green}{\textbullet} Detailed after 

guidance   & \textcolor{yellow}{\textbullet} Limited explanation       & \textcolor{red}{\textbullet} Missed key details         \\ \hline
\textbf{Practical Examples}     & \textcolor{green}{\textbullet} Provided examples         & \textcolor{yellow}{\textbullet} Theoretical only          & \textcolor{red}{\textbullet} No examples                \\ \hline
\textbf{Handling of} 

\textbf{Changes}    & \textcolor{green}{\textbullet} Understood after 

guidance & \textcolor{yellow}{\textbullet} Noted but shallow         & \textcolor{red}{\textbullet} Missed key changes         \\ \hline
\textbf{Depth of}

\textbf{Understanding} & \textcolor{green}{\textbullet} Deep with guidance        & \textcolor{yellow}{\textbullet} Moderate, 

limited depth   & \textcolor{red}{\textbullet} Lacked understanding       \\ \hline
\end{tabular}
\caption{Comparison of Document Handling by Different Models}
\label{tab:model_comparison}
\vspace{-1cm}
\end{table}


We demonstrate the learning capabilities of different models through an example that illustrates their ability to acquire new knowledge and explain it effectively. The \texttt{doacross} clause, introduced in the OpenMP 5.2 specification, is used with the \texttt{ordered} directive. Although \texttt{doacross} concept was mentioned in version 5.1, it was not formally specified as a clause until version 5.2. 
We tasked three LLMs with analyzing content from the OpenMP specifications about the \texttt{ordered} directive and then generating descriptions of the \texttt{ordered} directive and its clauses. Ideally, they would ascertain that the \texttt{doacross} clause is now the only clause used with the \texttt{ordered} directive and understand how to use it.
The primary objective of this study was to evaluate the LLMs' ability to extract and understand detailed information, with a specific focus on their comprehension of the \texttt{doacross} clause. 
Additionally, significant changes in the \texttt{ordered} directive have led to the removal of the use of \texttt{simd}, \texttt{threads}, and \texttt{depend} clauses, making \texttt{doacross} the sole clause used with \texttt{ordered} directive. Our analysis evaluated whether the LLMs could effectively learn from input files and not merely rely on pre-existing knowledge.

By analyzing the generated content, we noted that Claude's responses primarily offered general explanations and reiterations of the specification of the \texttt{doacross} concept. However, Claude did not provide specific insights into the \texttt{doacross} clause, such as its usage, functionalities, or applications. We believe that it failed to identify information specifically about the \texttt{doacross} clause, instead finding descriptions of the \texttt{doacross} concept in a different chapter. In other words, Claude does not seem to understand that \texttt{doacross} has been defined as a clause. 
Conversely, Gemini explained that the \texttt{doacross} is defined as a clause and this clause is used to specify explicit dependencies between iterations in parallel-executed loops. It highlighted that this enables the compiler to ensure the correctness of data dependencies as defined by developers, which is particularly crucial in scenarios with complex dependencies.

ChatGPT initially did not retrieve any content about the \texttt{doacross} clause when directed to study the \texttt{ordered} directive and its clauses. Instead, it used the \texttt{depend} clause incorrectly. However, when explicitly directed to focus on the \texttt{doacross} clause, ChatGPT demonstrated a deeper understanding and provided a more comprehensive response.  It accurately recognized that \texttt{doacross} was defined as a clause, explained its use, and even described its compatibility with the \texttt{ordered} directive: the \texttt{ordered} directive can be effectively combined with the \texttt{doacross} loop scheduling to provide finer control over dependencies in loop iterations. The \texttt{doacross} clause allows specifying dependencies between loop iterations, which is crucial for ensuring that iteration \(i\) completes specific tasks before iteration \(i+1\) can commence. It also showcased, through example code, how to use the \texttt{ordered} and \texttt{doacross} clauses to manage dependencies, applicable in scenarios requiring tightly coupled iterative operations.

The varying levels of understanding among the three LLMs concerning the newly added \texttt{doacross} clause present an interesting topic. Initially, we verified that all three LLMs could learn from new input materials and were not solely dependent on existing knowledge. ChatGPT does not process documents in detail, tends to overlook some content, and largely relies on existing knowledge bases. Yet, it can effectively understand and explain these details when specifically directed to extract particular content. Gemini shows strong abilities in processing the input materials; it was the only model to recognize \texttt{doacross}  as a clause without being explicitly prompted to focus on \texttt{doacross} clause. However, its explanations remain limited, resembling more a rephrasing of the specification content without offering an in-depth explanation or practical demonstrations of usage. Claude reads the input but fails to understand it correctly. It identifies content related to \texttt{doacross}  but does not recognize it as a clause nor grasp its application. It is unclear whether this is due to a misunderstanding or incorrect information retrieval.

Determining how much LLMs rely on user-uploaded content versus pre-existing knowledge is challenging.  This issue is a key area of interest in LLM algorithm design. In our work, we focus on evaluating the understanding capabilities of these models and the quality of the content they generate, without emphasizing the source of the information.

\subsection{Assessment of the \texttt{Interactive OpenMP Programming} Books}
\subsubsection{Quantitative Analysis of Generated Textbook}
In our published version of the \texttt{Interactive OpenMP Programming} book, there are over 200 OpenMP examples and more than 17,000 lines of text or code. We check the generated content for the text segments for accuracy and depth of understanding. This involves guiding the LLMs through several revisions or having different LLMs revise the content. We resort to manual edits if the ideal output is not achieved after multiple attempts. Typically, it takes a few minutes to generate the initial version of a chapter, followed by several hours of revisions. Completing the same volume of work totally manually usually requires over a week to finish a chapter's initial draft and revisions.

The code examples generated by ChatGPT and Gemini have been validated. They differ entirely from the official OpenMP examples, demonstrating that the LLMs do not simply copy content from input files. However, about 60\% of the complicated examples generated by Claude are directly from official examples from our experience. Most simple cases are completely accurate and do not require modifications, but more complex examples often need manual corrections or optimizations, with over 70\%.

\subsubsection{Official OpenMP Examples Versus Interactive Book Demonstrations}
We compared the examples using the \texttt{SIMD} directive in the \texttt{Interactive OpenMP Programming} book and the official OpenMP examples in Figure~\ref{fig:code_comparison}. The official example is a function that processes two double-precision arrays using \texttt{SIMD} directives for parallel reduction, emphasizing modularity in a potentially larger application. It uses a private temporary variable to ensure thread safety during the sum computation. In contrast, the example in our book is a complete standalone program within a main function, utilizing a single float array and demonstrating OpenMP's SIMD capabilities in a straightforward educational format. It includes array initialization, directly adds each element to the sum, and outputs the result, making it highly accessible for learning. Unlike the official example, which returns the computed sum for use elsewhere, the example in our book prints the sum directly, emphasizing immediate visual feedback for learners. 

\begin{figure} \scriptsize
     \centering
     \begin{subfigure}[b]{0.45\textwidth}
\begin{lstlisting}[language=C]
#include <stdio.h>
#define N 1024

int main() {
  float a[N]; float sum = 0.0f;
  for (int i=0; i<N; i++) a[i] = i*1.0f;
  // Vectorize the loop
  #pragma omp simd reduction(+:sum)
  for (int i=0; i<N; i++) sum += a[i];
  printf("Sum:%f\n", sum);
  return 0;
}
\end{lstlisting}
         \caption{Code Example of SIMD in Interactive OpenMP Programming Book}
         \label{fig:subfigure1}
     \end{subfigure}
     \hfill
     \begin{subfigure}[b]{0.49\textwidth}
        \begin{lstlisting}[language=C]
double work(double *a, double *b, int n) {
  int i;
  double tmp, sum;
  sum = 0.0;
  #pragma omp simd private(tmp) \
      reduction(+:sum)
  for (i = 0; i < n; i++) {
    tmp = a[i] + b[i];
    sum += tmp;
  } 
  return sum;
}
\end{lstlisting}
         \caption{Code Example of SIMD from Official OpenMP Examples}
         \label{fig:subfigure2}
     \end{subfigure}
\caption{Comparative Analysis of Code Examples}
\label{fig:code_comparison}
\vspace{-1cm}
\end{figure}

\section{Related Work}
\label{sec:related_work}
\subsection{Existing Books for OpenMP Programming}


In this section, we discuss notable contributions to OpenMP programming education, highlighting their relevance to our current study while identifying unique features, commonalities, and limitations.

"Parallel Programming in OpenMP" is one of the pioneering texts~\cite{chandra2001parallel}. Although it offers a historical and technical foundation, enriching our study's depth, the content has become significantly outdated, limiting its contemporary relevance. "Using OpenMP – Portable Shared Memory Parallel Programming" offers a comprehensive introduction to OpenMP, addressing hardware developments and comparing OpenMP with other programming interfaces~\cite{chapman2007using}. "OpenMP Common Core: Making OpenMP Simple Again" simplifies OpenMP by focusing on its twenty-one essential components~\cite{mattson2019openmp}. These books provide foundational knowledge and are rich in details and examples. However, they predominantly focus on basics and lack extensive real-world applications, with most missing in-depth optimization discussions.

"Using OpenMP – The Next Step" explores OpenMP's advanced features, such as tasking, thread affinity, and accelerators, delving into complex programming scenarios~\cite{van2017using}. "High-Performance Parallel Runtimes" provides an in-depth analysis of parallel programming models suitable for modern high-performance multi-core processors, with detailed discussions on optimizing key algorithms~\cite{klemm2021high}. "Programming Your GPU with OpenMP" focuses on GPU programming with OpenMP, emphasizing heterogeneous programming and performance optimization~\cite{deakin249programming}. These books cover advanced features in OpenMP's new standards and discuss deep optimization techniques, particularly hardware-specific optimizations, including CPUs and GPUs. While they offer valuable insights for optimizing performance, their complexity may pose challenges for beginners.


A collective examination of these OpenMP programming books reveals that they adhere to educational principles, structuring content from simple to complex, gradually deepening understanding. They emphasize practical learning through extensive examples, which are crucial for learners to grasp complex concepts practically. However, common limitations include rapid obsolescence due to the fast-paced evolution of parallel programming technologies, the inability to provide real-time feedback to users, and the long development cycle of traditional publishing, which cannot quickly address limitations once published.

Our \texttt{Interactive OpenMP Programming} book leverages LLMs like Gemini Pro 1.5, Claude 3, and ChatGPT-4 to quickly generate up-to-date content, enabling real-time code execution and a dynamic learning experience via Jupyter books. This "learning by practice" approach provides immediate feedback and practical application, making the material more engaging and customizable to individual learning needs. Additionally, the \texttt{Interactive OpenMP Programming} book is open-sourced and accessible online, reducing costs and increasing accessibility compared to traditional textbooks, which often require purchase and can be less accessible.

\subsection{The Use of LLM in Education and Textbook Writing}
LLMs like ChatGPT have demonstrated potential in aiding various stages of writing, including organizing material, drafting, and proofreading. One of the major challenges highlighted in educational settings is the accuracy and relevance of the information provided by the LLM, particularly how it integrates into the developers' workflow without disrupting it.
Xiao et al.'s study examines the utilization of ChatGPT for generating personalized reading comprehension exercises for middle school English learners in China~\cite{xiao2023evaluating}. Their research addresses the challenge of outdated and non-engaging educational materials by deploying ChatGPT to produce tailored reading passages and questions, thereby enhancing student engagement and material relevance. Through automated and manual evaluations, the system demonstrated its ability to generate educational content that often surpassed the quality of traditional methods. 
In a similar vein,
Nam et al. (2024) explore the potential of using LLMs to enhance code understanding and development within an Integrated Development Environment (IDE) through their prototype tool, GILT~\cite{nam2024using}. This tool integrates directly into the IDE to provide context-aware, real-time information support, aiming to help developers understand and expand unfamiliar code more effectively.
Their study involved a user study with 32 participants, showing that using GILT significantly improves task completion rates compared to traditional web searches.  
However, no significant gains were found regarding time savings or deeper understanding, suggesting areas for further improvement.

Another example concerning the relevance and accuracy of responses from LLMs comes from the work of Arora et al. (2024), who analyzed the usage patterns of LLMs among undergraduate and graduate students in advanced computing courses at an Indian university~\cite{arora2024analyzing}. Their research focused on how students employ LLMs for programming assignments, particularly in code generation, debugging, and conceptual understanding. Employing a mixed-method approach, combining surveys and interviews, the study highlighted that while LLMs significantly enhance student productivity by generating boilerplate code and aiding in debugging, they also present challenges regarding response accuracy and integration with student-generated code. This necessitates substantial student interaction with LLMs to integrate and troubleshoot system components effectively.
The findings emphasize the role of LLMs as supplementary tools in educational settings, suggesting the importance of proper prompts to enhance the utility and accuracy of LLM outputs in complex academic tasks. People should be very careful when verifying the accuracy of the generated code when using LLMs.

Altmäe et al. explore the use of ChatGPT in scientific writing, particularly focusing on its application in drafting a manuscript for reproductive medicine~\cite{altmae2023artificial}. The study illustrates the potential and challenges of using AI in academic writing, highlighting ChatGPT's role in streamlining content creation, manuscripts' initial composition, and refinement. Key challenges noted include the accuracy and relevance of AI-generated content, requiring significant human oversight to ensure scientific integrity, and raising ethical concerns about authorship and the potential for AI to discourage deep learning. This exploration aligns with broader discussions on integrating AI in educational tools, as seen in other research focused on programming education. It suggests AI's utility as a supplementary aid in complex intellectual tasks, provided its limitations are carefully managed.

Literature surveys are fundamental in academia and education, providing essential overviews of existing research and identifying future research directions. The study by Wang et al. introduces \texttt{AutoSurvey}, an innovative system designed to automate the creation of comprehensive literature surveys. Employing a systematic approach that encompasses initial retrieval and outline generation, subsection drafting by specialized LLMs, integration, refinement, and rigorous evaluation, \texttt{AutoSurvey} adeptly addresses the challenges posed by the vast volume and complexity of information. By leveraging the capabilities of LLMs, the system not only enhances the efficiency and quality of literature surveys but also demonstrates significant improvements in both citation and content quality compared to traditional methods. This exploration not only highlights the potential of LLMs to drastically reduce the time required to produce high-quality academic surveys but also underscores ongoing challenges such as context window limitations and the reliability of parametric knowledge within these models~\cite{wang2024autosurvey}.

The commonality among the studies above lies in their utilization of the powerful reading, understanding, and text-generation capabilities of LLMs. This aligns closely with our work. However, our research specifically focuses on generating OpenMP textbooks, emphasizing the teaching of OpenMP programming. We deeply explore the abilities of various LLMs to understand and interpret different OpenMP structures, parallel computing logics, and the generation of OpenMP examples. Additionally, we discuss how to continuously leverage LLMs to rapidly update textbook content, thereby addressing the critical issue of textbooks lagging behind during updates in OpenMP.




\section{Conclusion}
\label{sec:conclusion}
In conclusion, this paper has explored the innovative use of Large Language Models (LLMs) such as Gemini Pro 1.5, Claude 3, and ChatGPT-4 for creating the \texttt{Interactive OpenMP Programming} book. Our research indicates that while LLMs significantly enhance the interactivity and dynamism of educational content, they must be strategically integrated with traditional educational methodologies to maintain the depth and accuracy essential for effective learning. The developed interactive book, facilitated by Jupyter Notebooks, stands out by enabling real-time code execution and feedback, which is a considerable advancement over static learning materials. The success of our approach demonstrates that LLMs can play a crucial role in modernizing educational practices, especially in complex technical domains like OpenMP programming.

Future research should focus on refining the integration of LLMs into educational frameworks, enhancing the accuracy of content through improved prompt design, and exploring the scalability of this approach across other programming languages and frameworks. We also recommend ongoing assessments of the pedagogical impact of these tools to ensure they meet educational standards and effectively support learners. By continuing to leverage cutting-edge AI technologies, educators can better prepare students for the evolving demands of the tech-driven world, making learning not only more interactive but also more attuned to the needs of contemporary students.

\section*{Acknowledgement}
This material is based upon work supported by the National Science Foundation under Grant No. 2001580 and 2015254. This work was also prepared by LLNL under Contract DE-AC52-07NA27344 (LLNL-CONF-867264) and supported by the U.S. Department of Energy, Office of Science, Office of Advanced Scientific Computing Research, Scientific Discovery through Advanced Computing (SciDAC) program. 
\bibliographystyle{splncs04}
\bibliography{reference}

\end{document}